\documentclass[useAMS,usenatbib]{mn2e}

\title[The Eccentric Accretion Disc of A0620-00]
  {The Eccentric Accretion Disc of the Black Hole A0620-00\thanks{This
  paper includes data gathered with the 6.5-m Magellan Telescopes
  located at Las Campanas Observatory, Chile.}}
\author[Neilsen, Steeghs, \& Vrtilek]
  {J.~Neilsen,$^{1}$\thanks{E-mail: jneilsen@cfa.harvard.edu (JN);
  dsteeghs@cfa.harvard.edu (DS); svrtilek@cfa.harvard.edu (SDV)}
  D.~Steeghs,$^{2,3}$ S.D.~Vrtilek$^{2}$\\
  $^{1}$Harvard University Department of Astronomy, 60 Garden Street,
  MS-10, Cambridge, MA 02138\\
  $^{2}$Smithsonian Astrophysical Observatory, 60 Garden Street,
  Cambridge, MA 02138\\
  $^{3}$Department of Physics, University of Warwick, Coventry CV4 7AL, UK}
\date{Released 2007 Xxxxx XX}

\pagerange{\pageref{firstpage}--\pageref{lastpage}} \pubyear{2007}

\usepackage{latexsym}
\usepackage{color}
\usepackage{graphicx}
\usepackage{fancyheadings}
\usepackage{lscape}

\begin{document}

\label{firstpage}

\maketitle

\begin{abstract}
We present spectroscopic observations of the quiescent black hole
binary A0620-00 with the the 6.5-m Magellan Clay telescope at Las Campanas
Observatory. We measure absorption-line radial velocities of the
secondary and make the most precise determination to date
($K_{2}=435.4\pm0.5$ km s$^{-1}$). By fitting the rotational broadening of
the secondary, we
refine the mass ratio to $q=0.060\pm0.004$; these results, combined with
the orbital period, imply a minimum mass for
the compact object of $3.10\pm0.04$ M$_{\sun}.$ Although
quiescence implies little accretion activity, we find that the disc
contributes $56\pm7$ per cent of the light in B and V, and is subject to
significant flickering. Doppler maps of the Balmer lines reveal bright
emission from the gas stream-disc impact point and unusual
crescent-shaped features. We also find that the disc centre of symmetry does
not coincide with the predicted black hole velocity. By comparison
with SPH simulations, we identify this source with an eccentric
disc. With high S/N, we pursue modulation tomography of
H$\alpha$ and find that the aforementioned bright regions are strongly
modulated at the orbital period. We interpret this modulation in the
context of disc precession, and discuss cases for the accretion disc
evolution. 
\end{abstract}


\section{Introduction}
A0620-00 (V616 Mon) is the prototype Soft X-ray Transient, a class of
low-mass binary stars which exhibit infrequent but intense X-ray
bursts (Gelino, Harrison, and Orosz, 2001). In 1975 it became the brightest
X-ray nova ever detected, at approximately 50 Crab \citep{Elvis75},
and it was the first nova to be identified with a black
hole primary (McClintock \& Remillard, 1986; hereafter MR86). MR86
measured an orbital period of 7.75 hr and a radial velocity
semiamplitude for the K-type secondary of 457 km s$^{-1}$, leading to a mass
function $f(M)=3.18$ M$_{\sun};$ estimates of $K_{2}$ and $f(M)$ have
decreased slightly since then (i.e. 433 km s$^{-1}$ and 3.09 M$_{\sun}$)
(Marsh, Robinson, \& Wood 1994, hereafter MRW94). Given this minimum
mass, it is likely that A0620-00 is a black hole. 

A substantial amount of work has gone into the analysis of A0620-00 in
the last twenty years, with particular emphasis on ellipsoidal
variations in the light curve and the contamination of the K-star flux by
light from the accretion disc. As yet, no real consensus has been
reached, mostly due to the complexity of the lightcurves. While
ellipsoidal variations are obvious, they are highly asymmetric
(Leibowitz, Hemar, \& Orio 1998); the origin of the asymmetry is
undetermined.  Modelling this lightcurve, \citet{Gelino01} determined in
inclination of 41$\pm 3\degr$, invoking starspots to explain the
asymmetries. Shahbaz, Naylor, and Charles (1994)
found a 90 per cent confidence interval of $i=$30--45$\degr$ given the
mass ratio of A0620-00, modelling their asymmetries with the bright
spot where the accretion stream hits the disc.

Lightcurve modelling is also complicated by the variability of
the disc itself. To quantify ellipsoidal variations, most
authors assume the disc to be constant, and justify the claim by
noting that A0620-00 is quiescent. They do not mention that estimates
of the disc contamination range from $<$3 per cent \citep{Gelino01}
to $\la50$ per cent (MR86). The contribution from this disc is
not only unclear, but apparently not constant. More than half of A0620-00's
58-year burst cycle has passed, and it is important to note that
quiescent does not mean inactive. We will argue that the variability
of the accretion disc cannot be neglected. In order to determine
definitively the mass of the compact object, it is very
important to understand the structure and variation of the
accretion disc. MRW94 made enormous progress towards this goal.
In 2004, Shahbaz et al. (hereafter S04) noticed signatures of an
eccentric disc not seen in previous Doppler maps, but lacked the phase
coverage to verify their hypothesis. 

Therefore, as follow-up to the work of MRW94 and S04, and as part of a
Doppler imaging survey of black hole and neutron star binaries, we
undertook phase-resolved optical spectroscopy of
A0620-00. In $\S 2$ we describe our observational methods and data
reduction. In $\S 3$ we measure the radial velocity of the secondary
star, the system mass ratio, and attempt to quantify flickering. In
$\S 4$ we present Doppler images of the accretion disc at several
wavelengths, investigate evidence for an eccentric disc, and report
results of modulation tomography of the
H$\alpha$ line. We discuss conclusions from the variability of the
disc and our Doppler maps in $\S 5.$

\section{OBSERVATIONS}

We observed A0620-00 with the Low-Dispersion Survey Spectrograph
(LDSS3) at the f/4 focus of the 6.5-m Clay
telescope at Las Campanas observatory on 2006 December
14--16. We acquired 48 spectra using the VPH Blue grism and a
long 0.75 arcsec slit. By shifting the slit 4$\degr$ redward, we were
able to observe H$\alpha$ with the superior resolution of the Blue grism
(2.3 \AA~$\equiv$ 130 km s$^{-1}$), covering 4250--7035 \AA. To minimize the
effects of atmospheric dispersion, we observed at parallactic
angle. Our exposure times ranged from 420 s to 1200 s, with an average
of 660 s (a total of 8.81 hours on the source). 

Each night we observed the flux standard HILT 600
with the same instrumental setup as A0620-00. On all nights the seeing
was generally comparable to our slit width, but occasionally spiked
as high as 1.8 arcsec due to wind, and slit losses prevented precise
flux calibration. On 2006 Dec 15 we also
observed two K3/K4 stars, HD 18298 and HD 7142, as velocity standards,
again with the same optical setup. As the secondary is constrained to
be later than K3V \cite{FR07}, we expect more accurate
results for HD 7142, which is listed as K3/K4III in the SIMBAD
database (HD 18298 is listed as K3IIICN). For each pointing, we obtained
comparison HeNeAr arc lamp spectra after every 3--7 spectra, depending
on the current exposure time. We list the observations in Table 1.

\setcounter{table}{0}
\begin{table}
\caption{Observation log}
\label{tab1}
\begin{tabular}{@{}cccc}
\hline
Source  & Date   & \# of Spectra   & T$_{\rm exp}$ (s)\\
\hline
A0620-00 & 2006 Dec 14 & 11 & 873 \\
A0620-00 & 2006 Dec 15 & 23 & 698 \\
A0620-00 & 2006 Dec 16 & 14 & 433 \\
HILT 600 & 2006 Dec 14 & 1 & 20 \\
HILT 600 & 2006 Dec 15 & 3 & 20 \\
HILT 600 & 2006 Dec 16 & 3 & 20 \\
HD 18298 & 2006 Dec 15 & 3 & 5 \\
HD 7142 & 2006 Dec 15 & 5 & 5 \\
\hline
\end{tabular}

\medskip T$_{\rm exp}$ is the average exposure time for the source.
\end{table}
We used standard IRAF routines for basic data reduction
(zero-subtraction, flat-fielding, and spectral extraction). We
extracted our spectra in multispec format, attempting \textit{apall}'s
optimal extraction using nominal LDSS3 
gain, readout noise, and full-well values. The routine also performs
standard extraction and generates error bars. A CCD defect running
across our spectra prevented reliable fits to the spatial profile for
optimal extraction, so we used the normally extracted spectra
instead. This choice did not significantly degrade our S/N, and should
have a negligible effect on our presented results. After wavelength
calibration, we passed the spectra and their errors to the software
package \small{MOLLY}\normalsize~for cleaning and analysis.

\section{ANALYSIS}
\subsection{The spectrum of A0620-00}

\setcounter{figure}{0}
\begin{figure}
\includegraphics[width=84 mm]{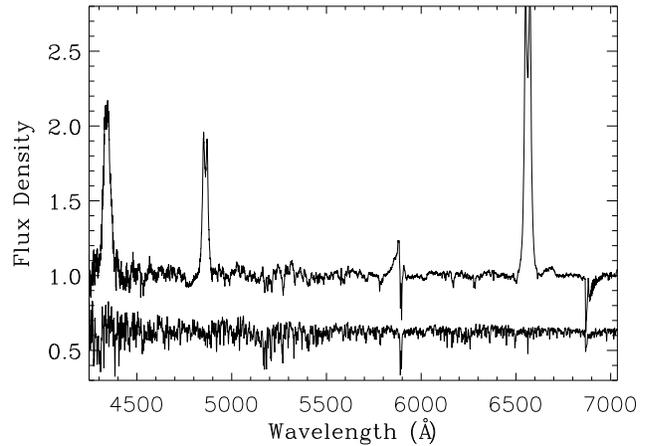}
\caption{The average normalized spectrum of A0620-00. The strongest
  features are, in order of increasing wavelength, H$\gamma,$
  H$\beta,$ He\,{\sc i} 5875, H$\alpha,$ He\,{\sc ii} 6678, and
  telluric absorption lines. We also show the scaled spectrum of HD
  7142 for reference.}  
\label{fig1}
\end{figure}

In Figure 1 we present the average normalized spectrum of A0620-00. The
spectrum shows a number of strong features (originating from the disc)
and many relatively weak K-star absorption lines. We overplot the K-dwarf
template HD 7142 for reference. We have
extremely high S/N H$\alpha$ and H$\beta$ lines, both observed by
MRW94, as well as lower S/N H$\gamma$ and He\,{\sc i} 6678\AA. The
non-detections of He\,{\sc ii} lines indicates the scarcity of
ionizing radiation. Our H$\alpha$ line is stronger relative to the
continuum than in 1994 by approximately 50 per cent. This relative
brightening is not surprising given the decreased fraction of light
contributed by the secondary. We show a close-up of H$\alpha$ in panel
a of Figure 2. The profile has two strong symmetric
peaks. It is also interesting that our
H$\beta$ line also shows two strong peaks (as opposed to a
single-peaked line seen by MRW94).

The other feature of note is located on top of the interstellar sodium
doublet near 5890 \AA. We show a close-up of this line in
panel b of Figure 2. The line suffers significant extinction by
interstellar sodium, so it is not possible to identify its peak
unequivocally, but it seems to be He\,{\sc i} 5875. It
appears to have some structure, and trailed spectra
suggest a double-peaked profile, but our attempts to correct for
interstellar absorption and create a Doppler map ($\S$ 4) were
unsuccessful. 

\setcounter{figure}{1}
\begin{figure}
\includegraphics[width=84 mm]{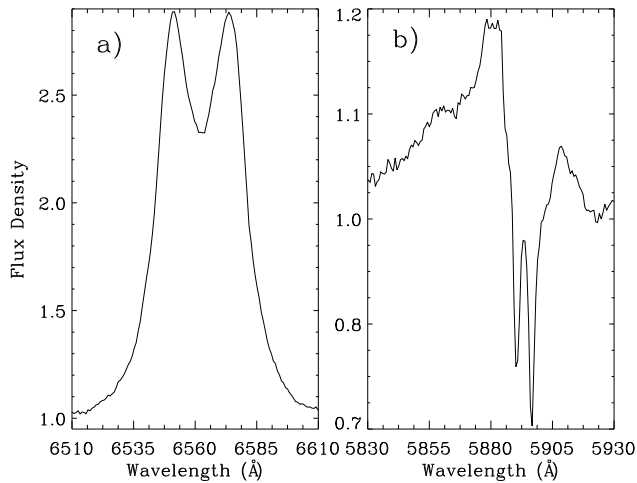}
\caption{Close-ups of a) the H$\alpha$ line and b) the feature near
  5890\AA. Intensities are as in Figure 1.}
\label{fig2}
\end{figure}

\subsection{The radial velocity of the secondary star}

Following MRW94, we first measure the radial velocity of V616 Mon in
order to estimate the mass function. Masking out emission lines,
telluric absorption, and
interstellar sodium lines in our velocity standards, and normalizing,
we cross-correlated our spectra against these K-dwarf templates. The
results of the following sections are presented in Table 2. 

\setcounter{figure}{2}
\begin{figure*}
\includegraphics[width=150 mm]{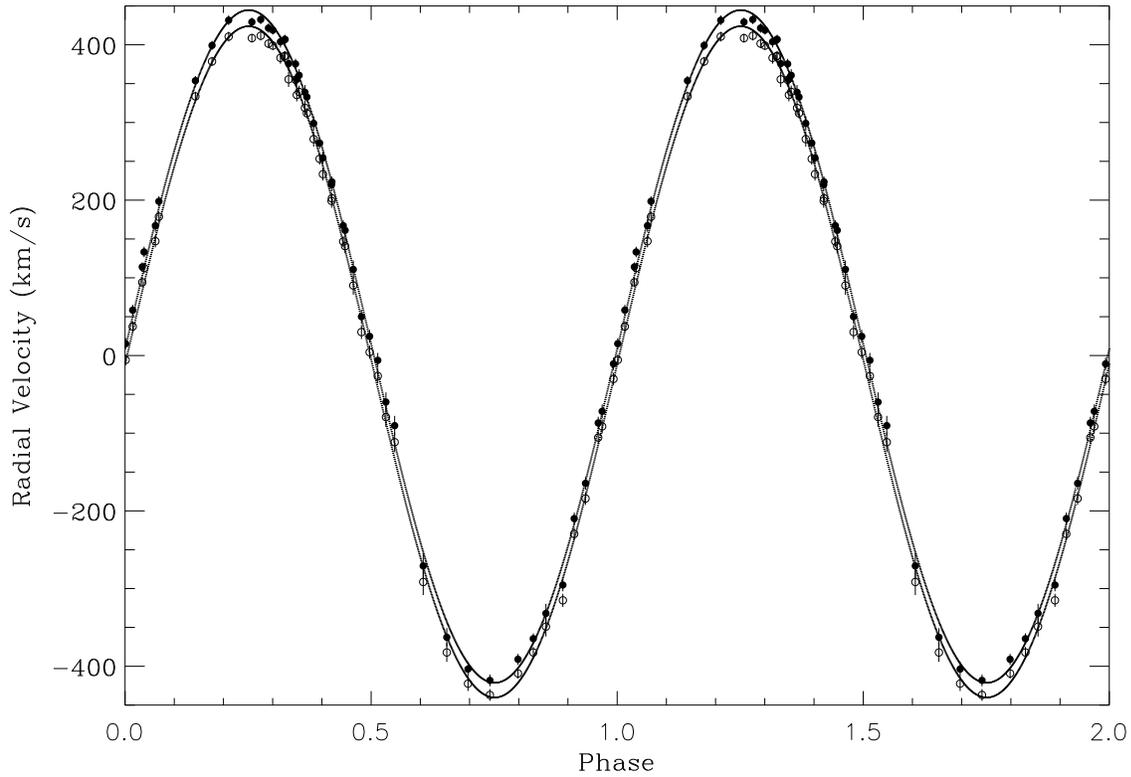}
\caption{Radial velocity of V616 Mon. Two cycles are shown
  for clarity. Filled circles correspond to the template HD 18298, and
  open circles to HD 7142. Note from Table 2 that the main difference
  is in $\gamma.$}
\label{fig3}
\end{figure*}

As discussed in MRW94, this process requires several adjustments for
A0620-00: rotational broadening and orbital smearing. The following
analysis was completed for each template. We performed a
preliminary broadening of the template with MRW94's value of $v\sin
i=83$ km s$^{-1}$. Then, after cross-correlating, we shifted our spectra into
the rest frame of the template, which is essentially the rest frame of
the secondary. We made 48 copies of the template and smeared each
according to the formula $s=2\pi V_{i} T_{i}/P,$ where $P$ is the orbital
period and $V_{i}$ and $T_{i}$ are the radial velocity and exposure
time for the $i$th A0620-00 spectrum. With exposure times of up to 1200
s, it is necessary to correct for smearing because $s$ is comparable
to $v\sin i.$ We averaged the smeared templates, rotationally broadened the
result with a value between 30 and 135 km s$^{-1}$, and optimally
subtracted the final template from the average object spectrum. We
assume a limb-darkening coefficient $\varepsilon$ of 0.65 (Wade \&
Rucinski 1985, and references therein), but perform these calculations
for $\varepsilon=0.45-0.85$ to evaluate our systematics. The
optimal subtraction routine returns the fraction $f$ of light
contributed by the secondary and a $\chi^{2}$ value. We fit
fourth-order polynomials to $\chi^{2}$ to identify the appropriate $f$
and $v\sin i.$ Results are included in Table 2.

The measured rotational broadening of V616 Mon ($80\pm2$ for HD 18298
and $83\pm2$ for HD 7142) is consistent with
MRW94, but $f$ is not, even though $f$ and $v\sin i$ were strongly
correlated (i.e. we were able to identify them with the same
$\chi^{2}$ minimum).  MRW94 found that the secondary contributes $\sim
94$ per cent of the light near H$\alpha$ and $\sim 85$ per cent of the
light near H$\beta.$ We observed both wavelengths simultaneously, and
find that the secondary contributes only 44 per cent of the light in
$B$ and $V$ (near 5500 \AA). This conclusion is effectively
independent of the limb-darkening coefficient. We will discuss
systematic uncertainties on $f$ in $\S3.4$.

Finally, we performed another cross-correlation of our spectra against
the template, this time with the appropriate value of $v\sin i.$ The
results are shown in Figure 3. Although we took arc exposures
frequently to minimize the effects of flexure, we found arc scales to drift
by $\sim$0.5 \AA~over the course of a few hours. As our dispersion
(0.69 \AA~pixel$^{-1}$ ) corresponds to 37.9 km s$^{-1}$ pixel$^{-1}$, it was
necessary to compensate for arc drift. We did this by
cross-correlating the telluric lines near 6900 \AA~and shifting out
the resulting velocities (generally around 10 km s$^{-1}$). In this way we
guaranteed a common (heliocentric) rest frame for all 48 spectra. 

We fit a function of the form $V=K_{2}\sin(2\pi
(t-t_{0})/P)+\gamma$ for each template. $\gamma$ represents the
systemic velocity in the template frame. Although we have very high
S/N, the baseline ($\sim$7 orbits) was insufficient
to make a reliable independent determination of the orbital period, so
we fixed $P$ according to the ephemeris of MR86. To get an
accurate measure of our statistical uncertainties, we performed 10,000
Monte Carlo simulations for each fit, assuming each radial velocity to be
distributed normally around the measured value, increasing the errors
from cross-correlation by a factor of three. For our final values,
we averaged the results from both templates and took the standard
deviation of the combined distribution as the uncertainty. The
resulting uncertainty is not purely statistical, as the use
of multiple templates includes some systematic errors
(e.g. template mismatch). These results can be found in Table 2. 

\setcounter{table}{1}
\begin{table*}
 \begin{minipage}{150mm}
 \caption{Fits to rotational broadening and radial velocities.}
 \label{tab2}
 \begin{tabular}{@{}cccccccccc}
\hline
Template  & $\varepsilon$ & $v\sin i$  & $f$ & $K_{2}$ (km s$^{-1}$)
& t$_{0}-2454084$ & $\gamma$ (km s$^{-1}$)   & $q$   & $f(M_{1})$
(M$_{\sun}$)   & $\chi^{2}$\\
 &  & (km s$^{-1}$) $\pm2$ & $\pm0.001$  & $\pm0.4$ & $\pm4$E-5 &
$\pm8$ & $\pm0.004$ & $\pm0.03$ &\\
\hline
HD 18298 &  0.45 & 77 & 0.441 & 435.8   &
0.69487  & 11  & 0.051& 3.06 & 1.06\\
~&  0.55 & 78 & 0.441 & 435.8   &
0.69487  & 11  & 0.052 & 3.07 & 1.06\\
~&  0.65 & 80 & 0.440 & 435.8   &
0.69487 & 11  & 0.057 & 3.09 & 1.05\\
~ &  0.75 & 80 & 0.441 & 435.8   &
0.69487  & 11  & 0.056 & 3.09 & 1.06\\
~ &  0.85 & 81 & 0.441 & 435.8   &
0.69487  & 11  & 0.058 & 3.10 & 1.06\\
HD 7142 &  0.45 & 80 & 0.433 & 435.0   &
0.69484  & -9  & 0.056 & 3.07 &1.03 \\
~ &  0.55 & 80 & 0.433 & 435.0   &
0.69484  & -9  & 0.057 & 3.08 & 1.03 \\
~ &  0.65 & 83 & 0.449 & 434.9   &
0.69484  & -9  & 0.063 & 3.11 & 1.02 \\
~ &  0.75 & 83 & 0.433 & 435.0   &
0.69484  & -9  & 0.061 & 3.10 & 1.03 \\
~ &  0.85 & 84 & 0.433 & 435.0   &
0.69484  & -9  & 0.064 & 3.12 & 1.03\\
\hline 
 \end{tabular}
 \end{minipage}
\end{table*}

All fits are comparable in quality and have
excellent $\chi^{2}$. We find consistent values for $K_{2}$ and
excellent agreement in $t_{0}$ -- we
observed inferior conjunction on the second night -- and our
measurements are independent of limb-darkening ($\varepsilon$=0.65 is
marginally preferred). We will consider the systemic velocities only
briefly. The uncertainties in $\gamma$ are dominated by the systematic
uncertainty (one fifth of a pixel, or $\sim$7.6 km s$^{-1}$). We were unable
to find a cataloged radial velocity for HD 18298, but from
\citet{Malaroda01}, the radial velocity of HD 7142 is 32.8 km
s$^{-1}$. Given that MRW94 report a systemic velocity of 22 km
s$^{-1}$, we consider these fits to be accurate; the choice of
template does not significantly affect $K_{2},~t_{0},$ or $v\sin i.$
In summary, we adopt $v\sin i=82\pm2$ and $K_{2}=435.4\pm0.5$ km
s$^{-1}.$ In addition, we find that inferior conjunction of the
  mass donor star, which defines the zero point $t_{0}$ of our
  ephemeris, occurs at HJD (UTC) 2454084.69485$\pm$0.00005. This
  latest $t_{0}$ is consistent with the ephemeris of MR86 (within 0.5
  $\sigma).$ 

\subsection{The mass ratio}

To measure the mass ratio, we use
Paczynski's (1971) formula relating the rotational broadening and
$K_{2}$ to $q$ for Roche lobe-filling stars,
\begin{equation} 
\frac{v\sin i}{K_{2}}=0.462[(1+q)^{2}q]^{1/3}.
\end{equation} 
This gives
$q=0.060\pm 0.004,$ where the uncertainty includes variation between
templates. This value is consistent with MRW94's measurement
(by the same method) of $0.064 \pm 0.01,$ and with their value of
$q=0.067\pm 0.01,$ obtained by calculating models on a grid over the
Roche lobe, including gravity darkening and quadratic limb
darkening. Although it might be possible with our improved resolution
to distinguish between the Paczynski approximation and the grid
models, since the grid model correction was far from
significant at their $1\sigma$ level, we opt to take our results as
accurate. This choice is validated by the weak (at best) dependence of
our measurements on limb darkening, and we report uncertainties large
enough to account for any systematics in $\varepsilon.$

Given the orbital period, $K_{2},$ and $q$, the minimum masses for
both objects are: 
\begin{eqnarray*}
f(M_{1})&=& 3.10\pm0.04~\rm{M}_{\sun}\\
f(M_{2})&=& 0.19\pm0.02~\rm{M}_{\sun},
\end{eqnarray*}
where the true mass goes like M$_{\rm min}/\sin^{3}i.$ As noted in many
papers (MRW94; Shahbaz et al. 1994), the mass of a maximally
rotating neutron star with the stiffest equation of state is 3.2
M$_{\sun};$ if causality is the only constraint, the absolute upper
limit is 3.76 M$_{\sun}$ \citep{FI87}. 
Using the constraints $39\degr\leq i \leq 75\degr$ \citep{Gelino01}, we find
3.4 M$_{\sun}\leq$ M$_{1}\leq 12.6$ M$_{\sun},$ with the most likely value
at 11.1 M$_{\sun}.$ We thus improve the precision of the mass function,
but it remains a possibility that A0620-00 is a stiff, massive neutron star.

\subsection{Ellipsoidal variations and the variability of the disc}

Part of the difficulty in measuring the parameters of A0620-00, and a
possible source of systematic error, is the tidal distortion of the
secondary. We assume, for example, in our radial velocity
measurements, that the secondary's centre of light coincides with its
centre of mass. For a star filling its Roche lobe with some
gravitational and limb darkening, this assumption is not likely to be
valid. Following \citet{Sterne41}, we checked for this distortion in
our data by fitting an extra $-K_{\rm ell}\sin(4\pi\phi)$ term to our
radial velocity curve. Orbital elements from \citet{Gelino01} lead
to a predicted $K_{\rm ell}=3.0$ km s$^{-1}$, but we found $K_{\rm
  ell}\sim 0.7$ km s$^{-1}$ (less than the error bars on each radial
velocity), and an insignificant improvement to $\chi^{2}.$ It is
possible that our data have insufficient S/N to make this
measurement. This particular effect is small, but we must be careful
not to dismiss the systematic uncertainties from ellipsoidal effects.


Ellipsoidal variations,
particularly in the lightcurve, receive a great deal of attention
because of the constraints they place on the orbital parameters,
especially the inclination. In practice, one usually attributes all variation
to the secondary (assuming the disc to be constant). However, the fact
that the fraction of light contributed by the disc changes by a
factor of nine over ten years casts significant doubt on that
particular assumption. In turn, this evolution must be accounted for
when interpreting lightcurves spanning a long period of time.

Given that we have good spectral resolution and relatively good phase
coverage, we have attempted to characterize the ellipsoidal
variations of the source during our observations. While it is most
common to address variations in the lightcurve, effects of tidal distortion
should also be apparent in the secondary light fraction
and the line equivalent widths. In the rest of this section we
quantify these variations.

\setcounter{figure}{3}
\begin{figure}
\includegraphics[width=84 mm]{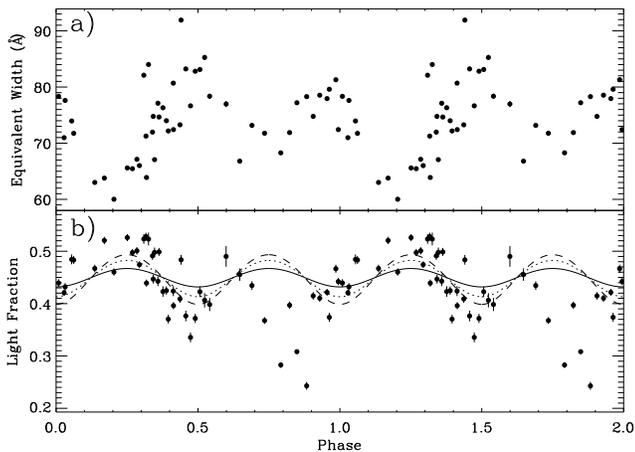}
\caption{a) Equivalent width of the H$\alpha$ line. b)
  Secondary light fraction as measured from 5000--5800 \AA~and
  6000--6500 \AA~for the
  template HD 7142. Differences from HD 18298 are less than 1$\sigma.$
  These plots indicate a flickering source of light. The solid,
  dotted, and dashed lines are the predicted light fraction for a
  constant disc and ellipsoidal modulations, for inclinations of
  40$\degr$, 60$\degr$, and 80$\degr$, respectively. See Equation 2.}
\label{fig4}
\end{figure}
If the disc is a constant diluting source of light, then continuum
variations, and therefore variations in the equivalent widths of
lines, maybe be attributed to the tidal distortion of the
secondary. In Figure 4a we show the equivalent width of the H$\alpha$
line as a function of orbital phase. The asymmetric ellipsoidal
variations observed by \citet{Gelino01} are obvious, along with
noticeable scatter. Given a S/N of $\sim 95$ at H$\alpha$, it seems
reasonable to interpret this as flickering, albeit undersampled,
rather than noise. We will discuss the equivalent widths in more
detail shortly.

Figure 4b shows the secondary light fraction
$f$ as a function of orbital phase. We can evaluate the variability of
the disc with a back-of-the-envelope calculation of $f(\phi).$ We
assume the secondary continues to exhibit ellipsoidal variations of
0.077 mag (MRW94 using $i=40\degr$), or flux variations of 7 per cent, and
that the disc is constant in time. Since the disc contributes
56 per cent of the light, it should be roughly 22 per cent brighter than the
secondary.  Then in units of the mean secondary flux,
 $F_{2}=1-0.07\cos(4\pi\phi)$ and $F_{1}=1.22,$ and 
\begin{equation}
  f=\frac{1-0.07\cos(4\pi\phi)}{2.22-0.07\cos(4\pi\phi)}.
\end{equation}  The inclination-dependent amplitude of
ellipsoidal variations is tabulated in MRW94. Equation 2 is plotted
along with $f$ for inclinations of 40, 60, and 80$\degr.$ We have
matched the mean relatively well, and to some extent the functional
form, but it is obvious that our assumptions do not hold. First, the
disc appears anomalously bright between $\phi=$0.7--0.9. As
this phase interval was observed on the second night only, we cannot
speculate if the dip in $f$ indicates a flare or a long-lived bright
region of the disc, like a warp, visible only at this phase.

The second and more troubling discrepancy is the large amplitude of
variation of $f.$ As is apparent from Figure 4b, excluding
$\phi=$0.7--0.9, higher inclinations are preferred, even those which
are ruled out by the lack of eclipses in this source. Accepting
momentarily the inclination determined by
\citet{Gelino01}, this figure illustrates very clearly the significance of the
assumption that the disc is a constant source of light; by requiring a
constant disc, we could overestimate the inclination by several tens
of degrees. Now it is reasonable to assume that the secondary has not evolved
substantially in the last twenty years, so we must conclude that the
discrepancy between the predicted and observed light fractions is
related to activity in the disc. For example, a component of disc
light modulating at the orbital period could reproduce the effect easily. 

Unfortunately, there is no a priori way to determine in advance the
viability of an inclination measurement. SMARTS data from the last ten
years show that A0620-00 goes through periods of quiescence, where ellipsoidal
modulations are observed, and periods of erratic variability which
gradually swamps the smooth component (Cantrell \& Bailyn 2007,
private communication). In these active periods, the source is
extremely variable on timescales from minutes (our observations) to
years (SMARTS), and the variability is highly
phase-dependent, so that ellipsoidal modulations cannot be reliably
measured; the current active state has persisted since December of 2003.
Associating this variability with the disc, we warn
against measurements of $i$ during such periods. We can expect,
furthermore, a strong correlation between flickering and the light
fraction. There is no reason $f$ cannot be small and constant
simultaneously, but an increase in light from the disc can only mean an
increase in accretion activity, for which flickering is highly
probable. 

This is admittedly a grim assessment of the situation, but lightcurve
estimates of the inclination made during periods of variability
or non-negligible contamination by disc light are unreliable. The
variation in $f$ itself is a final source of systematic
uncertainty. We have a very precise measurement of the dilution fraction
for the average spectrum, but that precision is only meaningful if $f$ is not
variable. Instead, we take the standard deviation in $f(\phi)$ as our
uncertainty: $f=44\pm7$ per cent, and summarize the discussion above
by reminding the reader that quiescence is not inactivity, and ought
not be used to justify invalid assumptions.

Consider now the equivalent width. Since the continuum varies
slowly over our lines, it is trivial to show that
the line equivalent width is given by \begin{equation}
  EW~(\textrm{\AA})\simeq\frac{EW_{\rm disc}}{1+f}, \end{equation}
where $EW_{\rm disc}$ is the ratio of flux integrated
over the line to the disc
continuum, and $f$ is the secondary light fraction. If the disc varies
uniformly or not at all, $EW_{\rm disc}$ should be a constant for any
given line. We can then attribute variations in the equivalent
width to variations in $f,$ which would be ellipsoidal in
nature. However, measurements of the equivalent width with the
secondary subtracted should reveal the reliability of this shaky assumption.

The equivalent width of our unsubtracted H$\alpha$ line (open circles)
is shown in Figure 5; it
should be compared to the same plot from MRW94. To facilitate this
comparison, we have overlaid their best-fitting line. We find good
agreement with their phasing and amplitude, but an increase in average
equivalent width of approximately 13 \AA. As noted earlier, the
H-alpha line is brighter relative to the continuum than it was in 1994, so
this increase is reasonable. Also shown in Figure 5 (filled
circles) is the equivalent width of the same line after subtraction of
the secondary. To achieve this result, which indicates the magnitude
of fluctuations in the disc, we normalized, broadened, and
smeared the template, subtracted $f(\phi)$ times the template from
each spectrum, and set the continuum to one. The dip between phases
0.7 and 0.9 corresponds to the possible flare seen in the light fraction.

We also note that the subtracted modulations are in phase with
the unsubtracted equivalent widths. As the secondary is expected to
contribute more light at longer wavelengths, the phasing may be an
artefact of improper secondary subtraction, because the strongest
contributions to $f$ come from 5000--5800 \AA. However, this effect should be
small because absorption lines up to 6500 \AA~were used in the
measurement. Even if this is the case, the scatter in $EW_{\rm disc}$
cannot be explained by the combined noise in $f$ and $EW.$ We suggest
that the most probable extra source of scatter is physical variability
of the disc, i.e. flickering. In this case, the fact that the
subtracted equivalent width is not constant suggests that the disc
fluctuations cannot be spatially uniform.  

\setcounter{figure}{4}
\begin{figure}
\includegraphics[width=84 mm]{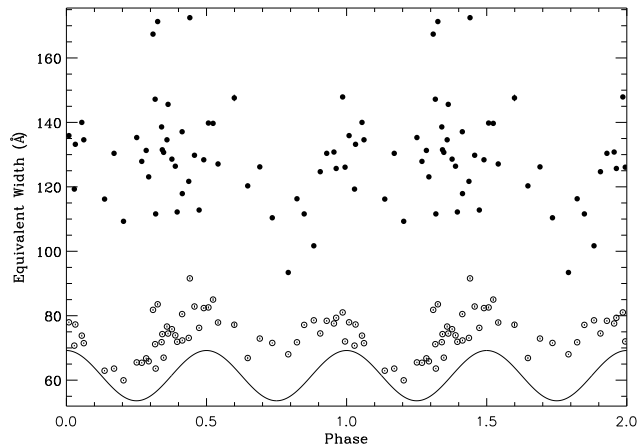}
\caption{Measured equivalent widths for the H$\alpha$ line from
  A0620-00. Open circles correspond to unsubtracted spectra; filled
  circles were calculated after subtracting the template HD
  7142. Again, differences between templates were insignificant, and
  the errors are smaller than the circles because of the extreme
  signal-to-noise. We have plotted MRW94's best fit for comparison.}
\label{fig5}
\end{figure}
Our measurements of flickering are confirmed by nearly
simultaneous observations with all four IRAC bands on Spitzer and the
1.2m FLWO telescope (McClintock 2007, private communication). These
observations, taken in the last week of November 2006, show strong erratic
variability which is well-correlated between telescopes. Neither
ellipsoidal variations nor the orbital period is obvious through
these fluctuations. The conclusion is clear: the
assumption that the disc is constant is not valid, though it may be a
good approximation if the disc contribution to the continuum is
negligible (for example, when \citet{Gelino01} found $f\ga 97$ per cent
and $i=41\pm 3\degr$). To investigate further the spatial
and temporal intensity of the disc, we present the results of Doppler
tomography and modulation tomography in $\S$ 4.

\section{Doppler Imaging}
In this section we discuss the results of Doppler tomography of the
emission lines from A0620-00. To create each image, we subtracted the
scaled HD 7142 spectrum, performed a linear
fit to the continuum around each line, normalized, and set the
surrounding continuum to zero (required by
\small{DOPPLER}\normalsize). MRW94 express some uncertainty as to the
propriety of interpolating over the H$\alpha$ line in the template
spectrum. Doing so removes an image of the donor star in the map,
and it is not clear which is the appropriate choice. However, we
found that Doppler maps including the light from the secondary do not
show donor emission, so we chose to interpolate over the line. Then we binned
the spectra to a uniform velocity scale and passed them to the
\small{DOPPLER}\normalsize~routine, which computes the maximum entropy
image that reproduces the observed line profiles during the course of
an orbital period. For an excellent summary of the method, see
\citet{Marsh01}.

\subsection{The maps}
In Figures 6--9 we present Doppler tomograms of the strong emission
lines in A0620-00. In each figure, the left column is, from top to
bottom: the observed trailed spectrum, the Doppler map, and
the fitted data. The right column is the observed data minus a
simulated symmetric part, the asymmetric part of the Doppler map, and
its fitted data. In the maps, we plot the Roche lobe of the secondary,
the ballistic trajectory of the gas stream (lower line), and the
Keplerian velocity of the disc along the stream (upper line).

In Figure 6 we show the maps of H$\alpha$. The most obvious feature is
the bright spot, which corresponds to the gas stream impact point. As
found by MRW94 and S04, in our maps the gas
stream trajectory and the Keplerian disc velocity along t he stream
(the two lines plotted in the map panels) bracket the bright spot. We
can interpret this as post-shock emission, originating
somewhat inside the outer edge of the disc \citep{Marsh90}. Using the
ballistic trajectory, we locate the spot at
$r=0.6\pm0.05R_{\rm L1}.$ If the disc velocities are Keplerian and the
spot moves with the disc, we find the outer edge near
$r=0.45\pm0.05R_{\rm L1}.$ This is some indication of our systematic
uncertainties, but we shall suggest shortly that the larger disc is
more likely. 

The other features of note are the two
crescents at $\sim 7$ o'clock and $\sim 2$ o'clock. We shall address
their origin shortly. The fitted data are rather
messy, mainly due to the presence of those features and the strong
flickering. Because the crescents and the bright
spot so thoroughly dominate the image, the symmetric part is too
bright, and we oversubtract to make the asymmetric part. Still, we do
detect some emission from the bright spot. The trail itself is not
particularly sinusoidal: it is more of a zigzag than an S-wave. The
discrepancy is best seen in the simulated asymmetric trail (bottom
right, Figure 6), but is apparent in the top panels as well. In the
data panel (top left), the trail is diagonal between $\phi=-3$ and
$\phi=-2.5$, but nearly horizontal at $\phi=-2$. It is difficult to
interpret this as the S-wave of a circular orbit. If the orbit is not
circular, it may be a  combination of the ballistic trajectory of the
stream and the motion of the disc, or it may be an elliptical orbit.

In Figure 7 we show maps of
H$\beta$. Again, we see the crescent features; their shape is more
apparent despite the lower S/N at H$\beta.$ For the same reason, we
have less trouble computing the symmetric part, and our asymmetric
part shows some emission along the stream. While the corresponding
trail does look nicer than its H$\alpha$ counterpart, the two seem to
be consistent. In Figure 8 we show maps of H$\gamma$, and in Figure 9 we
show maps of He\,{\sc i} 6678. These two maps, at much
lower S/N, reveal bright spot emission with some contribution from the
stream itself. 

\setcounter{figure}{5}
\begin{figure*}
\includegraphics[width=150 mm]{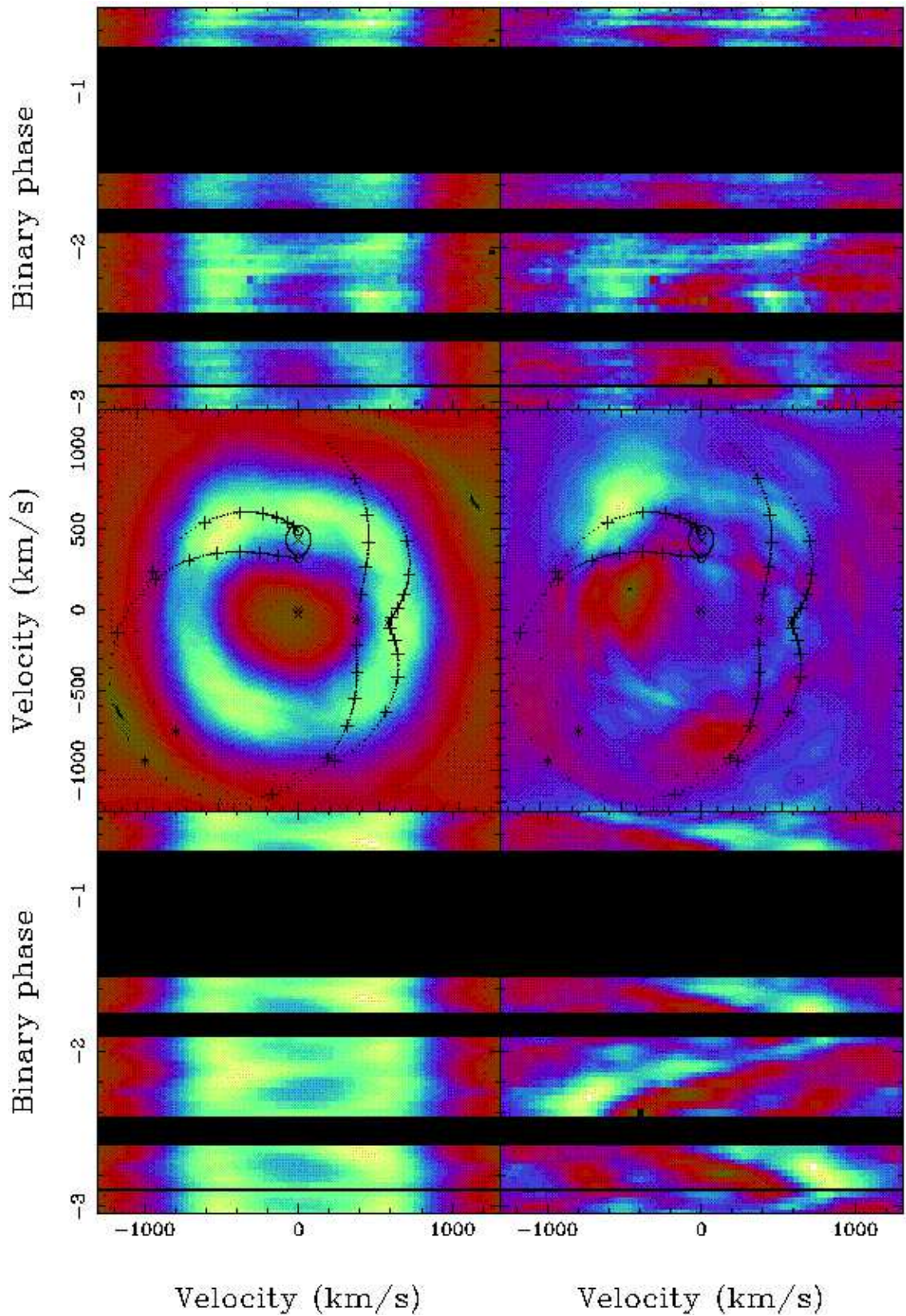}
\caption{H$\alpha$ Doppler maps. Notice in the middle left panel that
  the bright spot is located between the gas stream trajectory and the
  Keplerian velocity of the disc along the stream, and the brighter
  crescent-shaped features.}
\label{fig6}
\end{figure*}
\setcounter{figure}{6}
\begin{figure*}
\includegraphics[width=150 mm]{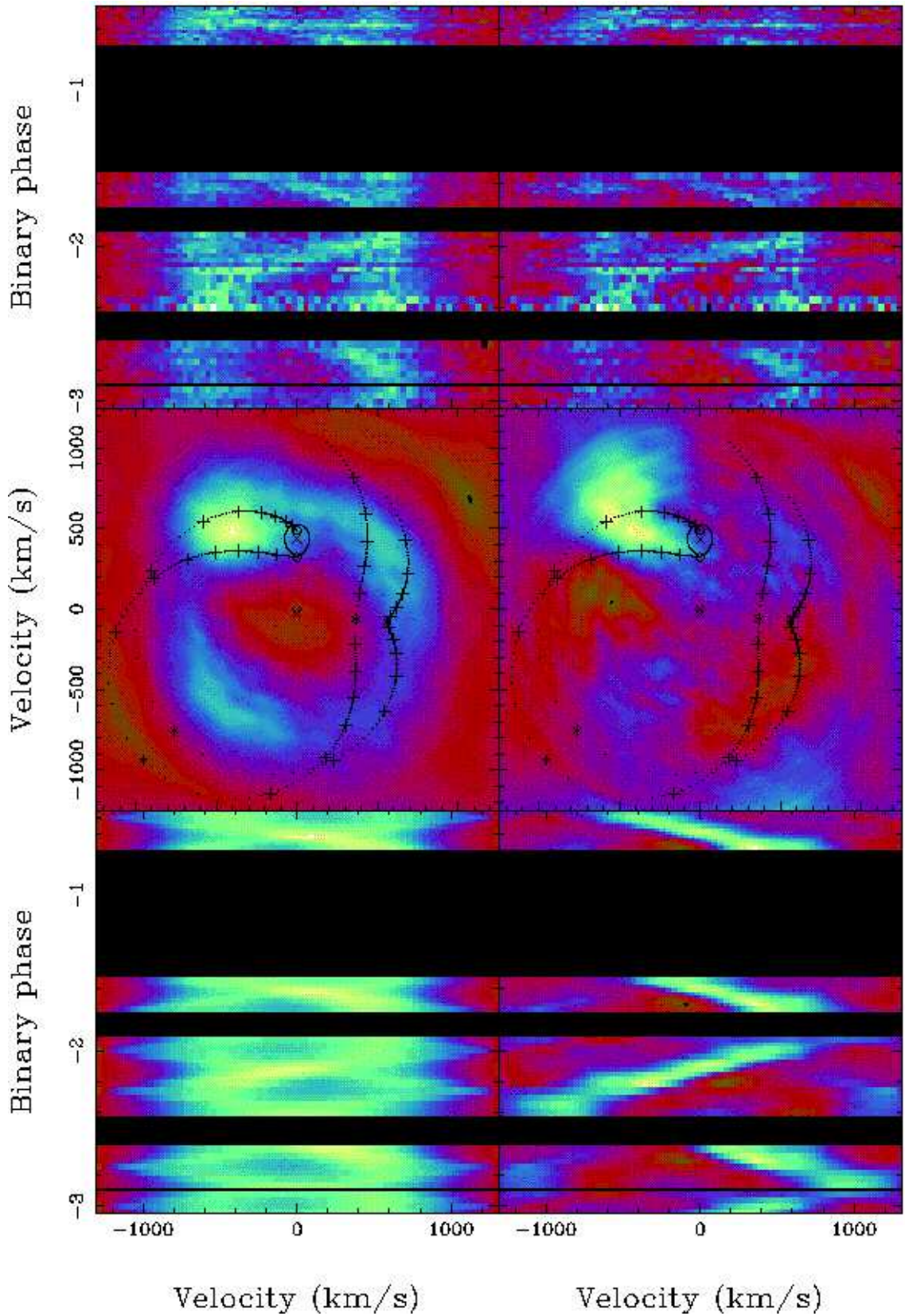}
\caption{H$\beta$ Doppler maps. The bright spot is again between the
  gas stream and the Keplerian disc velocity, and the crescents are
  obvious. The gas stream itself is clearly visible in the asymmetric
  part (middle right).}
\label{fig7}
\end{figure*}
\setcounter{figure}{7}
\begin{figure*}
\includegraphics[width=150 mm]{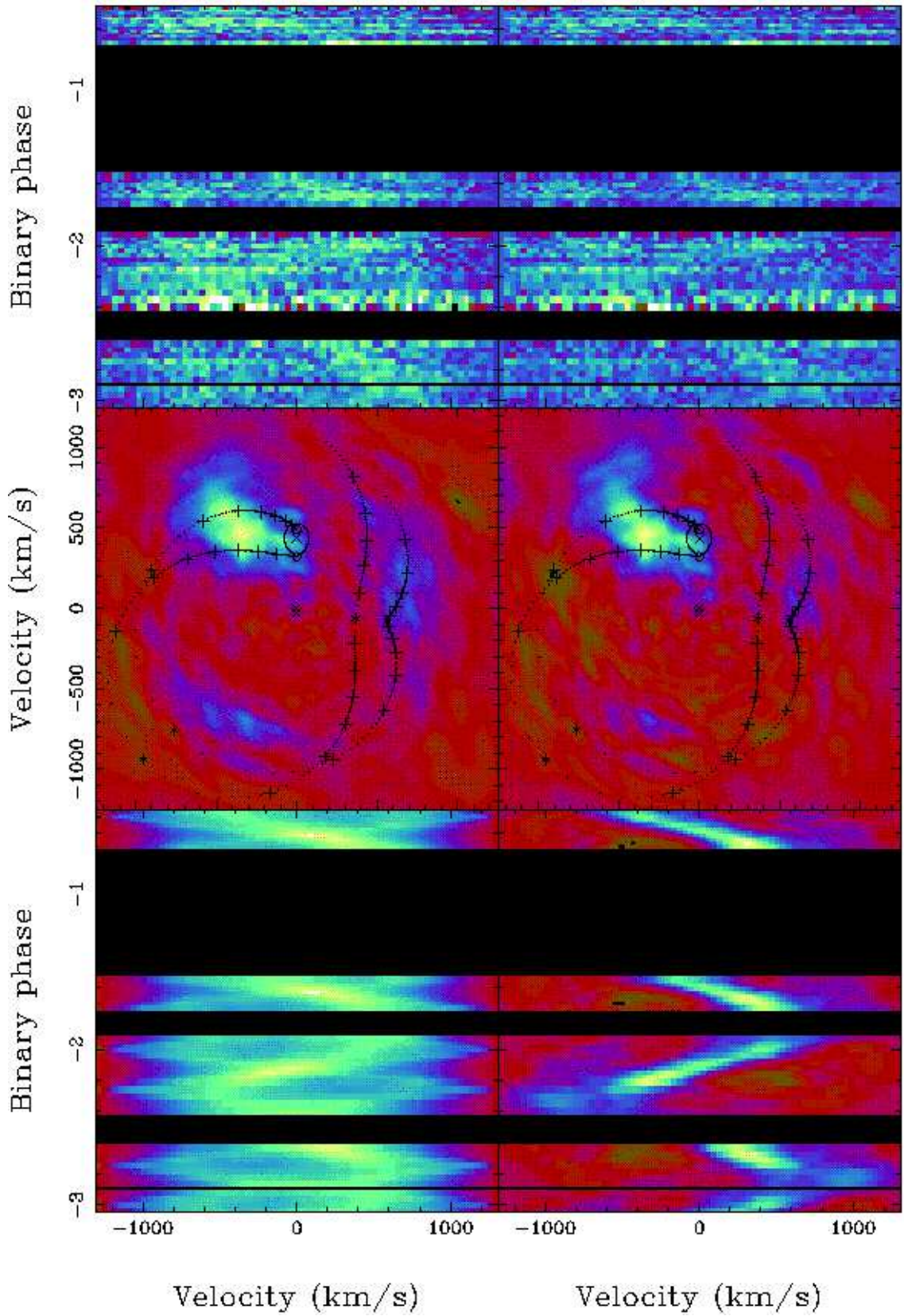}
\caption{H$\gamma$ Doppler maps. Only the bright spot is prevalent.}
\label{fig8}
\end{figure*}
\setcounter{figure}{8}
\begin{figure*}
\includegraphics[width=150 mm]{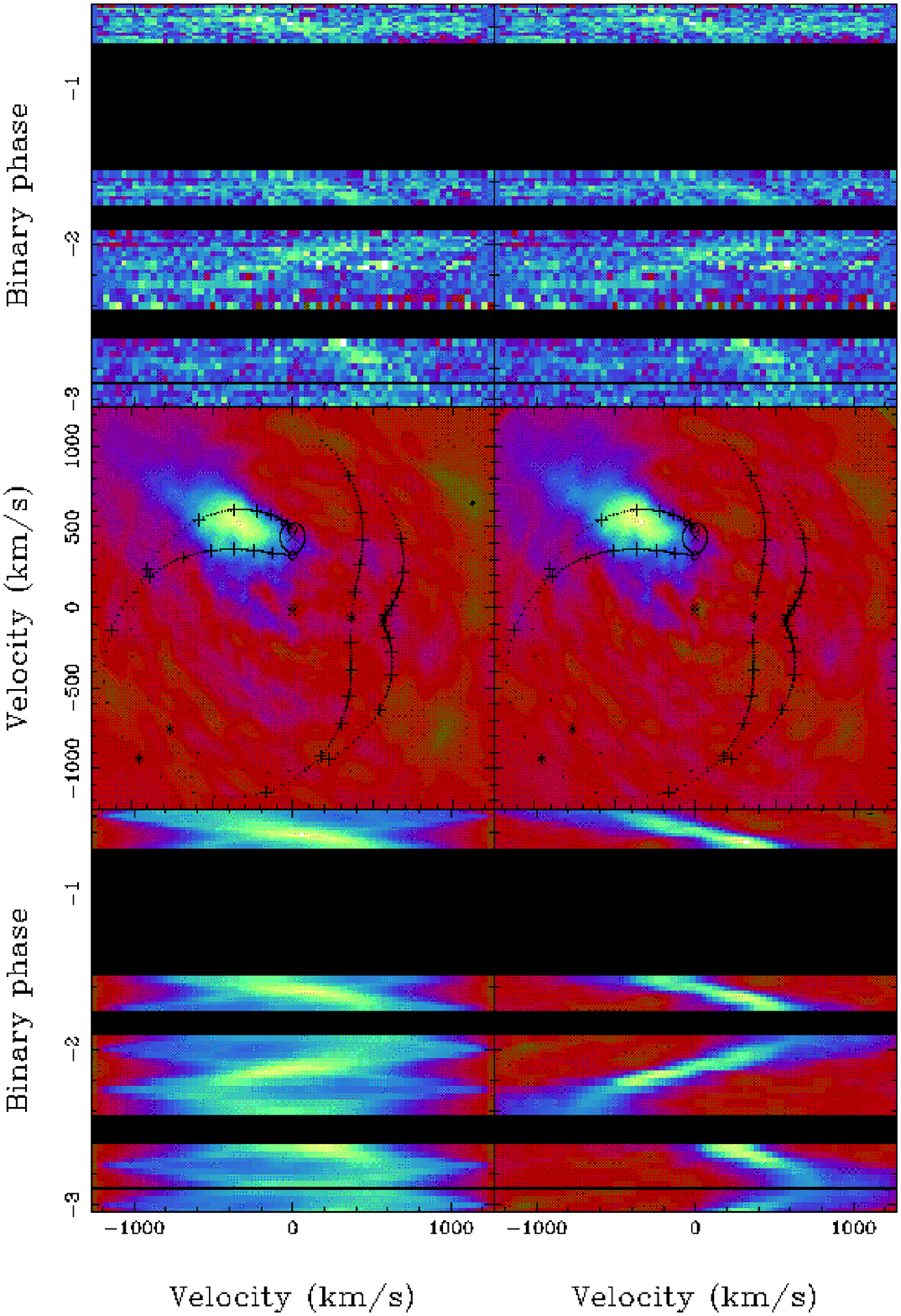}
\caption{He\,{\sc I} 6678 Doppler maps. Again, only the bright spot is
apparent.}
\label{fig9}
\end{figure*}

\subsection{The eccentric disc of A0620-00}
In the previous sections, we have presented evidence for a bright
flickering disc which contributes more than half of the light from the
source in the optical. It is already clear that the ``no intrinsic
variability'' assumption of Doppler tomography is not satisfied by
A0620-00. We should therefore exercise caution in interpreting our
Doppler maps. Tomography is a robust technique, but the physical
relevance of its results depends on the extent to which its
assumptions are violated. For example, we assume that all velocity
vectors corotate with the binary. However, if the disc is large
enough, it may reach an orbital resonance with the secondary and
become eccentric by tidal distortions. It will then precess, even in
the corotating frame. For the mass ratio of A0620-00, the dominant
resonance is 3:1 \citep{WK91}. Recent SPH simulations by
\citet{Foulkes04} show that the manifestations of eccentric discs
are bright emission between the gas stream and the Keplerian disc
velocity along the stream, non-sinusoidal S-waves, crescent-shaped
features in Doppler maps, and the shifting of the map centre of
symmetry away from (0,-$K_{1}$).

S04 observed the first three features and invoked an eccentric disc,
but they also point out one possible objection: MRW94 found a clean circular
disc. However, we have exposure times comparable to MRW94 with a
larger telescope; our higher signal to noise may have enabled us to
detect this phenomenon. The other explanation is that
the crescent was absent in 1994, and the source has changed. It is
thought that A0620-00 has an outburst recurrence time of 58 years, and more
than half of that time has passed since 1975, so the system should be
gearing up for a new outburst. In the disc instability model of
outbursts, the disc steadily recharges between outbursts, growing in
size and density until it becomes unstable. In the context of this
model, it is quite possible that since 1994, we
have actually watched the disc expand towards the 3:1 resonance
(r=0.66$R_{\rm L1}$) and become distorted. If this resonance is inside
the disc, then the distortion time-scale is $q^{-2}P_{\rm orb},$ or
about ninety days \citep{FKR02}. If not, the process is slower,
but a change over twelve years for A0620-00 seems reasonable. 

Now if the disc is actually circular, the Doppler map should be
radially symmetric about the point (0,-$K_{1}$), and by locating the
centre of the disc, we can identify the radial velocity of the black
hole \citep{Steeghs02}. Otherwise, the eccentricity of the
disc should be evident in a discrepancy between the observed and
predicted locations of this point. We therefore
implemented a search for the disc centre of symmetry, starting at the
predicted point and extending $\pm$200 km s$^{-1}$ in $V_{\rm x}$ and
$V_{\rm y},$ subtracting the symmetric part, squaring, and computing
the mean and standard deviation. The point with the lowest mean
corresponds to the centre of symmetry. We iterated our search,
updating the centre and improving the resolution until we found the
minimum. 

Since our disc is very structured, it was difficult to find a region
of the image unaffected by bright spots, so we  performed our search on
a smoothed version of the map. We then ran 10,000 Monte Carlo
simulations to find the centre of symmetry, using the standard
deviation of the residuals as the uncertainty to be sampled. Combining
the results for our H$\alpha$ and H$\beta$ maps, we find the center of
symmetry at (80,-220)$\pm$(40,20) km s$^{-1}$, well outside the
uncertainty in $K_{1}.$ Since we cannot explain a factor of nine of implied
increase in $q,$ we take this result, coupled with the bright spot
locationa and the crescent features in the maps, as evidence for the
non-zero eccentricity of the disc. 

Finally, we recognize that as the disc precesses, its apparent centre
of symmetry should move, as this point roughly corresponds to the mean
radial velocity of the disc. We made Doppler images of the first night
and the two halves of the second night, and performed the above search
for the centre of symmetry for each. However, the low phase coverage
of these maps results in large uncertainties, and we are unable to
determine a trend. The feasibility of such a measurement also depends on
the portion of the disc participating in eccentricity and precession,
and the exact precession period, which is generally estimated at $\sim
30$ orbits. With improved spectral resolution, a sequence of 3--4 full
nights (for sufficient phase coverage per night) sampling the
precession period should allow the motion of the center of symmetry of
the disc to be resolved.

\subsection{Modulation tomography}
It is also possible to relax the assumption of Doppler tomography that
the source flux is constant throughout the orbit \citep{Steeghs03}. The
new technique of modulation tomography allows not only the imaging of
average line emission from an accretion disc, but also maps harmonic
variations on the orbital period. The technique is robust and
flexible, but requires somewhat better S/N than standard tomographic
imaging. Therefore we only consider our H$\alpha$ profile here. The
process of creating modulation maps is quite similar to standard
tomography, and is described in \citet{Steeghs03}. The map is presented
in Figure 10.

\setcounter{figure}{9}
\begin{figure*}
\includegraphics[width=150 mm]{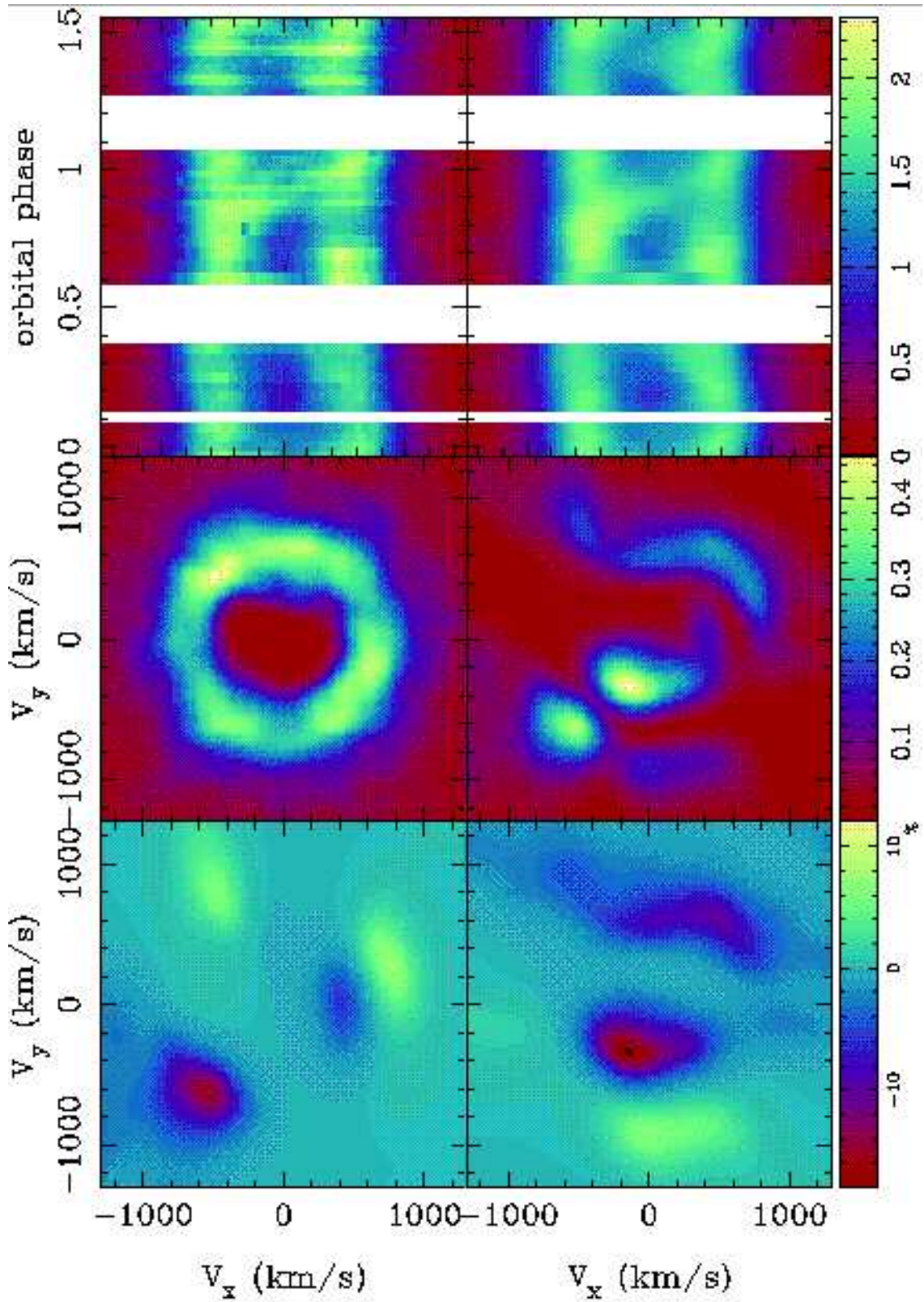}
\caption{H$\alpha$ modulation map. The observed data (top left) are
  well-reproduced by the fitted data (top right). The middle left
  panel shows the constant part of the disc, and the middle right
  panel shows the amplitude of modulation. The bottom row, left and
  right, are the cosine and sine components of variation,
  respectively. The crescent regions are obviously modulated at the
  orbital period.}
\label{fig10}
\end{figure*}
It is clear from the fitted data (top right), that modulation
tomography does a much better job reproducing the trail than standard
Doppler mapping. Whereas standard mapping could only reach
$\chi^{2}=35,$ we were able to attain $\chi^{2}=9.75$ with modulation
tomography. The poor $\chi^{2}$ is due to flickering. The constant
portion of the map (middle row, left
panel) looks quite similar to the maps presented in previous sections,
though the bright spot is significantly smaller. We still place it off
the gas stream trajectory, but it is much less diffuse, and is
located, as before, near $r=0.6R_{\rm L1}.$ It is clear from this image
that the inner edge of the disc in velocity space, and thus the outer
edge of the disc in physical space, extends as far as 0.7$R_{\rm L1}.$ The
crescents are not nearly as prevalent in the constant part, but are
apparent in the modulation maps. The crescent placed at 7 o'clock in
the standard map appears in the $\cos\phi$ map (bottom left) and the
crescent placed at 2 o'clock appears in the $\sin\phi$ map (bottom
right). The full modulating amplitude is shown in the middle row,
right panel.

For an erratic source like A0620-00, these maps must be interpreted
carefully. Modulation tomography, as mentioned, maps only harmonic
variations, and the demonstrated flickering is hardly
harmonic. However, we have shown the eccentric disc to be a viable
explanation for the observed phenomena, and the regions responsible
for the crescent emission are not constant by any means. It is clear
from the constant image
that the inner edge of the disc in velocity space, and thus the outer
edge of the disc in physical space, extends as far as 0.7$R_{\rm L1},$
lending credence to the 3:1 resonance argument. For typical
disc precession, superhumps are observed with periods $3-5$ per cent off the
orbital period \citep{WK91}. If indeed this modulating
emission is caused by viscous dissipation at the outer edge of an
eccentric precessing disc, there should be some power at $P_{\rm orb},$
though a map of variation at some superhump period $P_{\rm sh}$ would
reveal larger amplitudes. As yet, such a map is not
possible. None the less, modulation tomography has provided yet another
piece of evidence for our eccentric precessing disc model, and the
fact that there is a component of disc light modulating at the orbital
period could explain the observed light fraction from the secondary.

\section{Discussion}

We have presented spectroscopic analysis of the black hole binary
A0620-00. We measure an absorption-line radial velocity
$K_{2}=435.4\pm0.5$ km s$^{-1}$. With two measurements of
the rotational broadening of the secondary, we find a mass ratio of
$q=0.060\pm0.004$ and a minimum mass of
$3.10\pm0.04$ M$_{\sun}$ for the primary object. With the most
likely inclination of 41$\degr$ from \citet{Gelino01}, measured in
$J,~H,$ and $K$, the black
hole has a mass of 11.1 M$_{\sun}.$ The strong infrared flickering
discussed earlier, in conjunction with unexplained smooth variability
in the lightcurve and uncertainty in the disc spectrum itself, makes
it difficult to estimate the true uncertainty in the inclination. The
range of reported inclinations, 31$\degr$ to 70.5$\degr$(Gelino et
al. 2001 and references therein), results in a black hole mass between
3.7 M$_{\sun}$ and 22.7 M$_{\sun}$. Until the nature and variability
of the light from the disc is revealed in full, this conservative
error estimate must be sufficient. 

We also find that the secondary contributes $44\pm7$ per cent of the light
near 5500 \AA. As this means that the disc contributes a significant
fraction of the light, especially in emission line regions, it becomes
important to assess the variability of the disc, particularly if the
inclination is to be determined by lightcurve modelling. As noted, it
is common to assume that the disc is a constant source of light. While
it may be valid when $f$ is large, three observational points
cast doubt on this assumption:
\begin{enumerate}
\item S04 performed a detailed study of flares from
  A0620-00, which in their observations have amplitudes nearing 20 per
  cent of the source flux.
\item Measurements of the fraction $f$ of light contributed by the
  secondary have not been consistent. MR86 found $40\pm10$ per cent at
  5100 \AA, MRW94 found $94\pm3$ per cent at H$\alpha$, and
  \citet{Gelino01} found $f\ga97$ per cent in $J,~H,$ and $K$,
  assuming a \textit{constant} diluting source of light.
\item Observations in all four Spitzer bands, taken approximately two
  weeks before our observations on the Clay telescope, show strong
  flickering which is highly correlated with simultaneous R-band
  lightcurves from the 1.2m telescope on Mt. Hopkins.
\end{enumerate} 
While McClintock, Horne, and Remillard (1995) rightly point out that the
absorption line strength of the template will affect the observed
dilution fraction, unless it can be shown that the measured fraction is
strongly correlated with template line strength, a physical origin for
this variation cannot be ruled out. Future studies could assess the
true dependence of $f$ on template star, as well as the long-term
variability of $f$, by observing both BS 753 (MWR94's template) and HD
7142. Our measurements of the H$\alpha$ equivalent
widths, when compared to those of MRW94, suggest a physically real
origin for the variation, because equivalent widths are independent of
the template star and the instrument. So there appears to be evidence
for a disc which is brighter relative to the secondary than it used to be. 

Indeed, the modulations of the equivalent width even point to a
physically non-uniform flickering, because the disc emission lines
vary relative to the non-stellar continuum. We can tentatively
identify the flickering with the crescents, neither of which was
present in 1994, so our conclusion seems viable. The line emission
and the continuum may have different radial emissivity
dependencies, which could result in slower modulations. We also detect
several bright regions in the disc: one near the gas stream impact
point, and two crescent-shaped regions on opposite sides of the
disc. The reality of these features, as well as the non-uniform
flickering, are confirmed by modulation tomography of the H$\alpha$
disc line, which reveals variation near the bright
crescents. 

First noticed by S04, the crescent-like features may
indicate an eccentric disc, which is predicted for systems like A0620-00
with small mass ratios and large discs. It seems that we have observed
an eccentric disc, but let us consider the evidence. From our
observations, the following are clear:
\begin{enumerate}
\item The disc is bright and variable. The brightness is evident in the
   increased dilution of the secondary spectrum, and the variability
   is clear from a number of phenomena. First, the trailed H$\alpha$
   line shows clear evidence of flickering events. Second, the
   subtracted equivalent width of the same line is variable beyond
   explanation by noise alone. Third, the phase-resolved light
   fraction cannot be reproduced by ellipsoidal variability on
   top of a constant source of light. 

\item The disc extends to the 3:1 tidal resonance. This is a simple
   point, clear from the Doppler and modulation maps.

\item The disc is not centered on the radial velocity of the black
   hole.

\item The disc is not radially symmetric, but characterized by bright
   crescent-shaped regions.

\item The crescent regions are modulated at the orbital period.
\end{enumerate}

Each of these points alone would be insufficient evidence to
conclude that the accretion disc is eccentric and
precessing. But with the exception of a direct image of the elliptical
disc, we can present a complete and coherent argument that this is
the case. The disc has grown to tidal resonance, where the enhanced
disc viscosity results in bright and variable rims of extra dissipation.
We then observe crescents of extra dissipation at relatively low
velocities, as expected. Given the viscous effects, it is predicted
that the disc will receive a gravitational torque from the secondary,
and begin to precess. The asymmetries introduced here shift the
velocity center of disc emission away from the black hole, and we find
that the disc is not centered on the black hole in velocity
space. Furthermore, given the beat period between the precession and
orbital motion, the regions of viscous dissipation should be modulated
at roughly the orbital period. Modulation tomography reveals this to
be the case. 

In retrospect, knowing that portions of the disc are modulated on the
orbital period, we look closer at the fraction of light contributed by
the secondary star, and see that it is not well fit by ellipsoidal
modulations for a system at the inclination of A0620. But if another
component of the system was variable on the orbital period, as we have
observed the disc to be, then there is no need for concern. The
physical picture, a precessing elliptical disc torqued by the
secondary star, predicts and produces all the phenomena we have
discussed in our data, which are of high quality.

To put it another way, the eccentric disc hypothesis is nicely
self-consistent. It explains why and how the accretion disc has
changed, allows the disc to be large enough for the growth of
eccentric modes, and predicts the phenomena that we observe. It is
unfortunately not possible at this point to make an estimate of the
disc eccentricity. \citet{Smith07} have shed a great deal of light on
the evolution of disc eccentricity and energy dissipation with 3D SPH
simulations. They find that systems with $q$ between 0.08 and 0.24
develop low-mass eccentric discs withsuperhumps; for $q=0.0526,$ the
disc exhibits a short-lived superhump and decaying eccentricity. All
mass ratios show enhanced dissipation in the disc from the
thermal-tidal instability, even without the eccentric modes. 

Since we have not observed a superhump, we cannot place A0620-00 in either
category. If it falls in the more extreme group, the disc eccentricity
is likely zero (reached after about 300 orbital periods)
\citep{Smith07}. In that case, the steady state is a massive disc. If the
steady state is very long-lived, and the disc continues to grow, this
could explain the enormous intensity of novae like A0620-00. If
it fits among
the less extreme mass ratios, the disc eccentricity is around 0.1--0.2, and a
superhump should be observable with better photometry and a longer
baseline \citep{Smith07}. A0620-00 may also be at a transition
between those cases, and its evolution might be somewhat more
erratic, as suggested by the SMARTS data discussed earlier. For
example, it may toggle between states of quiescence,
superhumps, and variability (like what we have observed here). It
might, then, be erroneous to interpret this recent increase in
brightness as the build towards outburst. 

While we have strong evidence that the accretion disc around the
black hole has grown out to the tidal distortion radius, evolved into
an eccentric disc, and started to precess, further study is required
to verify our conclusion. Data from SMARTS, FLWO, and Spitzer will
further quantify flickering, and may reveal a superhump, or some new
period consistent with our results, and future programs of tomography
will track the evolution of the accretion disc. In anticipation of the
impending outburst, and in light of progress in simulations, we
suggest that this well-studied system not be
disregarded or ignored, for it affords us the opportunity to watch the
evolution of an accretion disc from quiescence to outburst, and the
chance to test models for disc instabilities in X-ray novae. 

\section*{Acknowledgements}This research was supported by the NSF
grant AST-0507637, the Harvard 
University Graduate School of Arts and Sciences (JN), and a SAO Clay
Fellowship (DS). We wish to thank Cara Rakowski for help with the
observations, Tom Marsh for use of his software packages, Jack
Steiner for many useful discussions, and the reviewer for a number of
constructive comments which improved the quality of the paper.

{}

\label{lastpage}

\end{document}